\documentclass[onecolumn,superscriptaddress,showpacs,pra]{revtex4}
\usepackage{amsfonts}
\usepackage{graphicx}   
\usepackage{graphics}
\usepackage{epsfig}

\usepackage{float} 

\usepackage{anysize}
\marginsize{3cm}{3cm}{3cm}{3cm}
\begin{document}
\title{A Split-Step Fourier Scheme for the Dissipative Kundu-Eckhaus Equation and Its Rogue Wave Dynamics}

\author{Cihan Bay\i nd\i r}
\email{cihanbayindir@gmail.com}
\affiliation{Associate Professor, Engineering Faculty, \.{I}stanbul Technical University, 34469 Maslak, \.{I}stanbul, Turkey. \\
						 Adjunct Professor, Engineering Faculty, Bo\u{g}azi\c{c}i University, 34342 Bebek, \.{I}stanbul, Turkey. \\
						 International Collaboration Board Member, CERN, CH-1211 Geneva 23, Switzerland.}

\author{Hazal Yurtbak}
\affiliation{Department of Civil Engineering, I\c s\i k  University, 34485 Maslak, \.{I}stanbul, Turkey}


\bigskip
\begin{abstract}
We investigate the rogue wave dynamics of the dissipative Kundu-Eckhaus equation. With this motivation, we propose a split-step Fourier scheme for its numerical solution. After testing the accuracy and stability of the scheme using an analytical solution as a benchmark problem, we analyze the chaotic wave fields generated by the modulation instability within the frame of the dissipative Kundu-Eckhaus equation. We discuss the effects of various parameters on rogue wave formation probability and we also discuss the role of dissipation on occurrences of such waves. 

\end{abstract}

\maketitle
\bigskip

\section{Introduction}

The Eckhaus equation is a nonlinear partial differential equation which is an extended version of the well-known nonlinear Schr\"odinger equation (NLSE). This equation was introduced by Kundu \cite{Kundu1984} and Eckhaus \cite{Eckhaus1985} independently, therefore it is commonly known as Kundu-Eckhaus equation (KEE). The KEE admits many different types of analytical solutions including but not limited to the single, dual and N-solitary waves, seed solutions and rogue wave solutions \cite{Levi2009, Wang2013, BayPRE1, BayPRE2}. KEE is used to model various phenomena such as fiber optical waveforms, water waves, fluids, ion-acoustic waves just to name a few \cite{Levi2009, Wang2013, BayPRE1, BayPRE2}. 

One of the most striking features of the nonlinear systems such as the KEE is their ability to sufficiently describe unexpectedly large waves. These waves, which are unexpected and have heights on the order of at least two times the significant wave height in a chaotic wave field, are known as rogue waves. Rogue waves appear in optics, hydrodynamics, plasmas and in finance \cite{BayPRE1, BayPRE2, BayEx_Kee, BayArxivKEE_self}.

The effect of losses or gain are taken into consideration in some nonlinear models i.e. the dissipative nonlinear Schrödinger equation \cite{Demiray2003}. However, to our best knowledge, such effects are not studied within the frame of the KEE before. With this motivation, we study the dissipative Kundu-Eckhaus equation (dKEE) in this paper. We first derive a simple analytical solution and then use that solution as a benchmark problem to analyze the stability and accuracy of a split-step scheme we propose for the numerical solution of the dKEE. We show that modulation instability leads to rogue wave formation within the frame of the dKEE. We discuss the effect of the dissipation parameter on the probability of occurrences of rogue waves.

\section{Methodology}
The dissipative Kundu-Eckhaus equation (dKEE) can be written as
\begin{equation}
i\frac{\partial U}{\partial t}+\mu_1 \frac{\partial^2 U}{\partial \xi^2}+{\mu_2} \left|U\right|^2 U+i{\mu_3}U + \mu_4^2\left|U\right|^4 U-2\mu_4 i\left(\left|U\right|^2\right)_{\xi}U = 0,
\label{eq2_1}
\end{equation}
where $t$ is the time and $\xi$ is the space parameter. In this equation, the parameter $\mu_1$ is the dispersion constant, the parameter $\mu_2$ is the cubic nonlinearity constant and the parameter $\mu_4$ is the quintic nonlinearity and Raman scattering constant. The parameter $\mu_3$ controls the dissipation or gain, depending on its sign \cite{Demiray2003}. Seeking a solution to the dKEE in the form of
\begin{equation}
U(\xi,t)= a\left(t\right)e^{i\left[k\xi-\Omega\left(t\right)\right]}
\label{eq2_2}
\end{equation}
one can obtain a simple solution as
\begin{equation}
U= Ae^{-\mu_3t}e^{i\left[k\xi-\mu_1k^{2}t-\frac{\mu_2}{2\mu_3}A^{2}e^{-2\mu_3t}- \frac{\mu_4^{2}A^{4}}{4\mu_3}e^{-4\mu_3t}+c\right]}
\label{eq2_3}
\end{equation} \\
where $A$ and $c$ are constants. We use this simple exponential solution as a benchmark problem to test the stability and accuracy of the split-step Fourier scheme we implement in the next section.

\subsection{A Split-Step Fourier Method for the Numerical Solution of the dKEE}

In this section we propose a split-step Fourier method (SSFM) for the numerical solution of the dKEE. As in the other spectral methods \cite{Pathria1990, Demiray2015, BaySR}, the SSFM calculates the spatial derivatives using FFT routines in periodic domains \cite{Dockery1996, Weideman1986, Ablowitz1984, Bogomolov2006, Sinkin2003, Stoffa1990, BayindirMak4}. However, temporal derivatives are calculated using a stepping procedure. In SSFM, the governing equation is splitted into two parts generally, namely the linear and nonlinear part. Various order splittings are possible for the utilization of the SSFM. As a possible first order splitting, we split the nonlinear part of the dKEE as
\begin{equation}
iU_t= -(\mu_2\left| U \right|^2 +  \mu_4^2 \left| U \right|^4-2i\mu_4 (\left| U \right|^2)_{\xi} + i\mu_3)U
\label{eq09}
\end{equation}
which can be integrated to give
\begin{equation}
\tilde{U}(\xi,t_0+\Delta t)=e^{i(\mu_2\left| U_0 \right|^2 +  \mu_4^2 \left| U_0 \right|^4-2i\mu_4 (\left| U_0 \right|^2)_{\xi} + i\mu_3 )\Delta t}\ U_0   
\label{eq10}
\end{equation}
where $\Delta t$ is the time step and $U_0=U(\xi,t_0)$ is the initial condition. One can evaluate the spatial derivate in this equation using the Fourier transforms
\begin{equation}
\tilde{U}(\xi,t_0+\Delta t)=e^{i \left(\mu_2\left| U_0 \right|^2 +  \mu_4^2 \left| U_0 \right|^4-2i\mu_4 F^{-1}\{ ikF[\left| U_0 \right|^2] \} +i\mu_3 \right)\Delta t}\ U_0   
\label{eq11}
\end{equation}
where $k$ is the Fourier transform parameter. In here, $F$ and $F^{-1}$ denote the forward and inverse Fourier transforms, respectively. All Fourier transforms are evaluated using efficient FFT routines in this study. The remaining linear part of the dKEE can be written as
\begin{equation}
iU_t=-\mu_1 U_{\xi\xi}
\label{eq12}
\end{equation}
Using the Fourier series one can evaluate the linear part as
 \begin{equation}
U(\xi,t_0+\Delta t)=F^{-1} \left[e^{-i\mu_1 k^2\Delta t}F[\tilde{U}(\xi,t_0+\Delta t) ] \right]
\label{eq13}
\end{equation}
where $k$ is as before. Therefore, pluging Eq.(\ref{eq11}) into Eq.(\ref{eq13}), the complete form of the SSFM for the numerical solution of the dKEE can be written as
 \begin{equation}
U(\xi,t_0+\Delta t)= F^{-1} \left[e^{-i\mu_1 k^2\Delta t} F[ e^{i(\mu_2\left| U_0 \right|^2 +  \mu_4^2 \left| U_0 \right|^4-2i\mu_4 F^{-1}[ikF[\left| U_0 \right|^2]] +i\mu_3  )\Delta t}\ U_0 ] \right]
\label{eq14}
\end{equation}
Throughout this study, the number of spectral components are selected as $N=1024$ and $\Delta t=10^{-4}$ which does not cause any instability in the SSFM simulations.

 \section{Results and Discussion}
\subsection{Comparisons of the Analytical and Numerical Solutions of the DKEE}
\begin{figure}[htb!]
\begin{center}
	\includegraphics[width=0.9\textwidth]{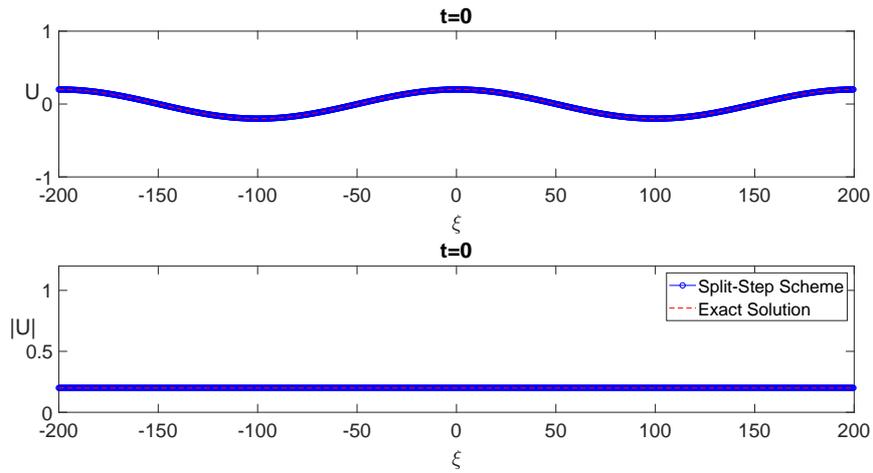}
  \end{center}
\caption{\small Comparison of the split-step vs exact solution of the dKEE at $t=0.0$ for  $\mu_1=1, \mu_2=2, \mu_3=0.1, \mu_4=2/3, A=0.2$.}
  \label{fig1}
\end{figure}

In this section, we provide a comparison of the analytical solution of the dKEE given by Eq.(\ref{eq2_2}) and its numerical solutions obtained using the SSFM. With this purpose, in Fig.~(\ref{fig1}), we compare the real part and absolute value of those complex valued solutions at $t=0$ for  $A=0.2, c=0, \mu_1=1$, $\mu_2=2$, $\mu_3=0.1, \mu_4=2/3$. 

\begin{figure}[htb!]
\begin{center}
	\includegraphics[width=0.9\textwidth]{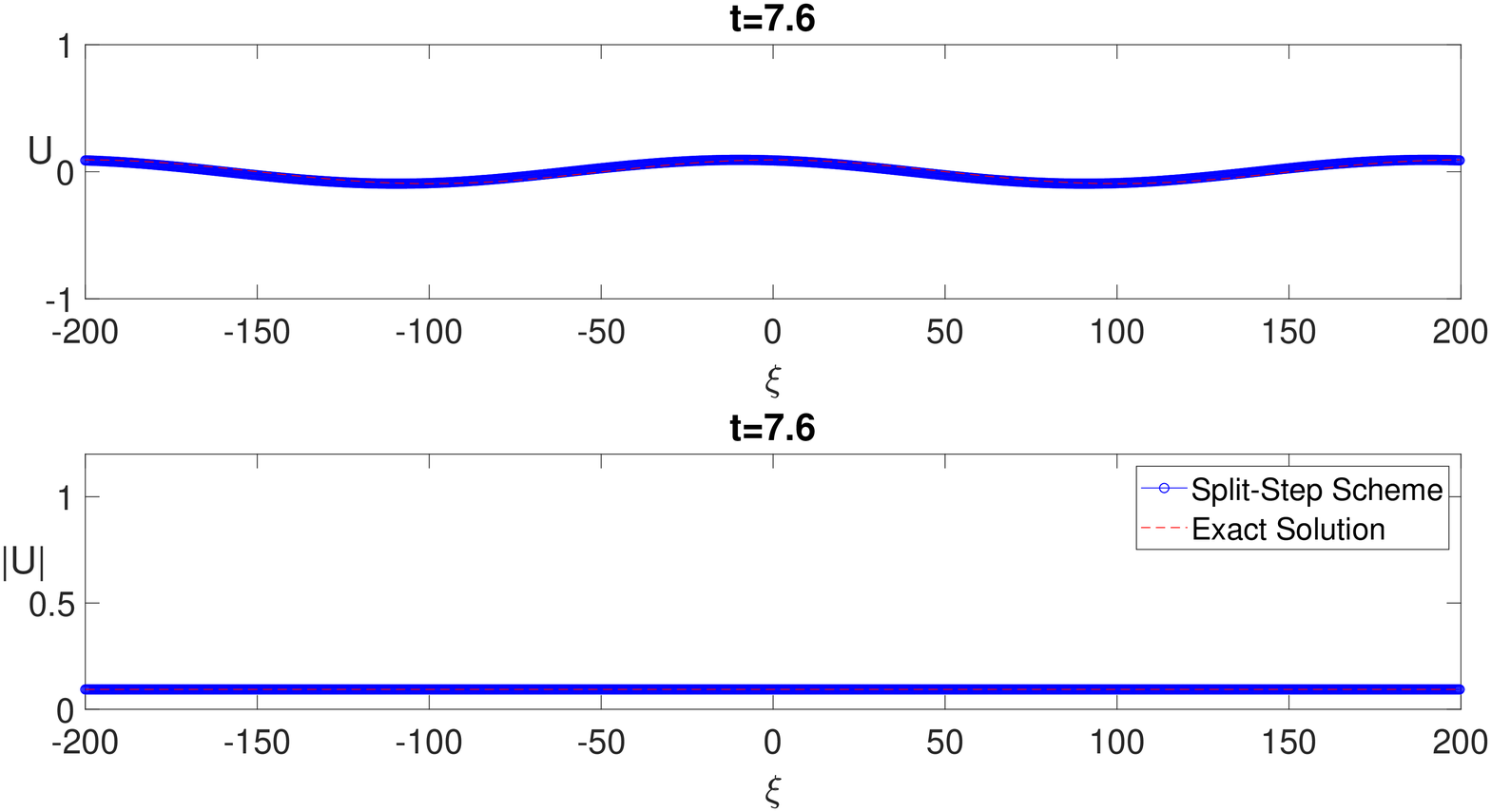}
  \end{center}
\caption{\small Comparison of the split-step vs exact solution of the dKEE at $t=7.6$ for $\mu_1=1, \mu_2=2, \mu_3=0.1, \mu_4=2/3, A=0.2$.}
  \label{fig2}
\end{figure}

\begin{figure}[htb!]
\begin{center}
	\includegraphics[width=0.9\textwidth]{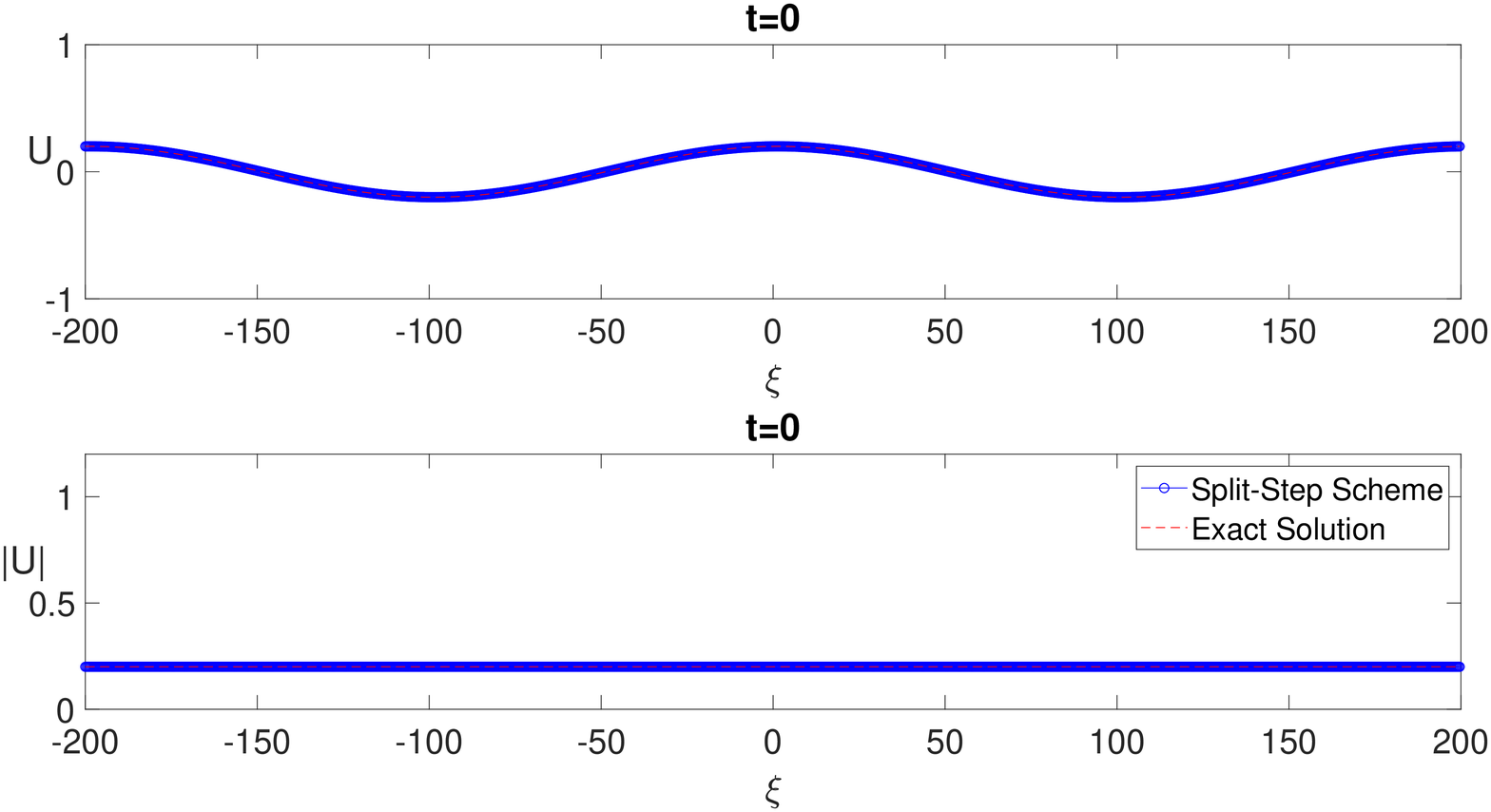}
  \end{center}
\caption{\small Comparison of the split-step vs exact solution of the dKEE at $t=0.0$ for $\mu_1=1, \mu_2=2, \mu_3=1, \mu_4=2/3, A=0.2$.}
  \label{fig3}
\end{figure}

\begin{figure}[htb!]
\begin{center}
	\includegraphics[width=0.9\textwidth]{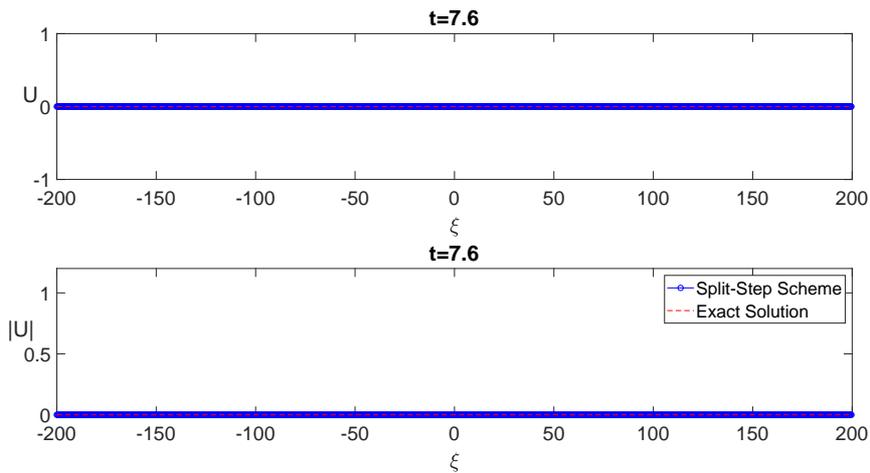}
  \end{center}
\caption{\small Comparison of the split-step vs exact solution of the dKEE at $t=7.6$ for $\mu_1=1, \mu_2=2, \mu_3=1, \mu_4=2/3, A=0.2$.}
  \label{fig4}
\end{figure}

As one can realize by checking the figure, the two solutions at the initial stage is in agreement. After time stepping is performed using the SSFM, the numerical and analytical solutions are still in good agreement at $t=7.6$, as depicted in Fig.~(\ref{fig2}). The effect of non-zero dissipation coefficient becomes significant after time stepping, the waves and the envelope of the wave field, which can be obtained by using the Hilbert transforming wavefield, tends to decrease as depicted in the Fig.~(\ref{fig2}).

Next, we turn our attention to the case where the dissipative effects are stronger. Changing the dissipation coefficient $\mu_3$, and selecting the same parameters as before, that is by setting $A=0.2, c=0, \mu_1=1, \mu_2=2, \mu_3=1, \mu_4=2/3$, we perform the numerical simulation again and plot the comparative results for $t=0$ in Fig.~(\ref{fig3}) and for $t=7.6$ in Fig.~(\ref{fig4}). As one can realize from these figures, the analytical and numerical solutions are in good agreement and the proposed SSFM for the numerical solution of the dKEE can be used safely. Additionally, by comparing Fig.~(\ref{fig2}) and Fig.~(\ref{fig4}), one can realize the significant effect of increasing the dissipation parameter, $\mu_3$. The value of $\mu_3=1$ imposes a very strong dissipation in the frame of the dKEE and the solutions decay within few dimensionless time units.

\subsection{Statistics of Rogue Waves of the DKEE and the Effect of Dissipation}

Rogue waves are considered as the unexpected and high amplitude waves. They are generally desired in fiber optical media, however their results can be catastrophic in the marine environment. There are some studies for their early detection \cite{BayindirPLA}. One of the triggering mechanisms that transforms sinusoidal wave trains into chaotic wave trains having abnormally high waves is the Benjamin-Feir instability. This instability is known as the Benjamin-Feir instability, or more commonly as the modulation instability (MI) \cite{Feir1967, Benjamin1967, Akhmediev1986, Zakharov2005, Zakharov2009}. In order to discuss the effects of dissipation on the rogue wave formation probability within the frame of the dKEE, we trigger MI in our numerical simulations. In order to trigger MI, a sinusoidal solution with a white noise is generally used as an initial condition. Therefore, in order to create random wave fields having rogue wave components, we use an initial condition for SSFM in the form of
\begin{equation}
U_0=e^{imk_0\xi}+ \beta a
\label{eq15}
\end{equation}
In here, $m$ is a constant, $k_0$ is the fundamental wave number which is equal to $2\pi/L$, $\beta$ is MI parameter and $a$ is a set of uniformly distributed random numbers in the interval of $[-1,1]$. Various values of $m$ and $\beta$ are considered in this study, which may lead to different probabilities of rogue wave occurrences.

\begin{figure}[htb!]
\begin{center}
	\includegraphics[width=0.9\textwidth]{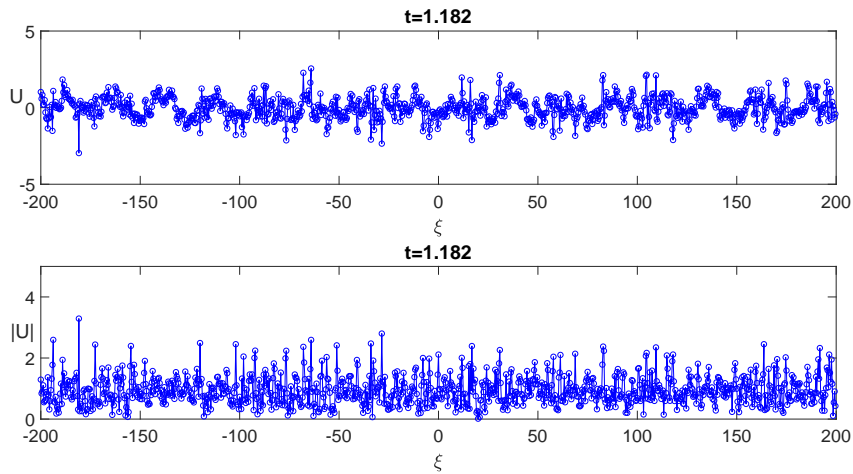}
  \end{center}
\caption{\small A typical chaotic wave field generated in the frame of dKEE for $m=16, \beta=0.4, \mu_1=1, \mu_2=2, \mu_3=0, \mu_4=2/3$.}
  \label{fig5}
\end{figure}
In Fig.~(\ref{fig5}), we depict a typical chaotic wave field exhibiting rogue wave components generated within the frame of dKEE. The parameters of computation are selected as $m=16, \beta=0.4, \mu_1=1, \mu_2=2, \mu_3=0, \mu_4=2/3$ for this simulation. It is useful to note that we start our simulations with a sinusoid having unit amplitude with a white noise superimposed on it and during time stepping we observe that waves having amplitudes of $\left|U \right|=0-5$ are occurring. The waves having amplitudes $\left|U \right|> 2$ can be classified as rogue waves for this simulation.

\begin{figure}[htb!]
\begin{center}
	\includegraphics[width=0.9\textwidth]{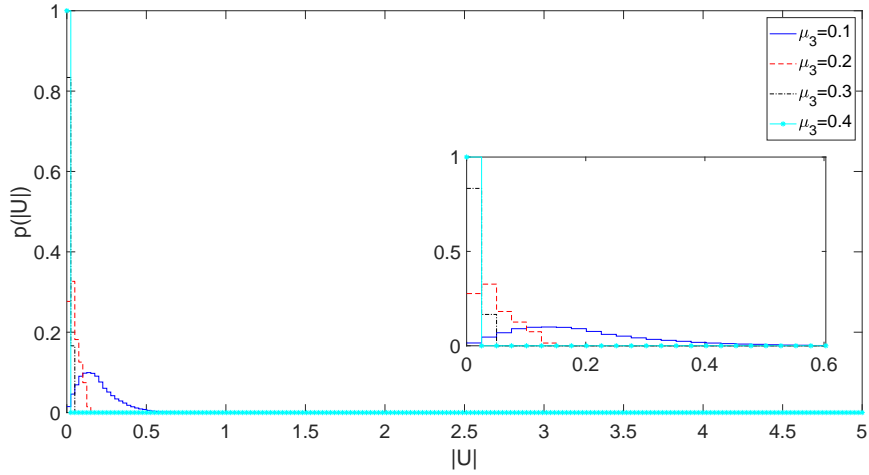}
  \end{center}
\caption{\small Amplitude probability distribution in a chaotic wave field for $m=4, \beta=0.1, \mu_1=1, \mu_2=2, \mu_4 = 2/3$ for various values of $\mu_3$.}
  \label{fig6}
\end{figure}
\begin{figure}[htb!]
\begin{center}
	\includegraphics[width=0.9\textwidth]{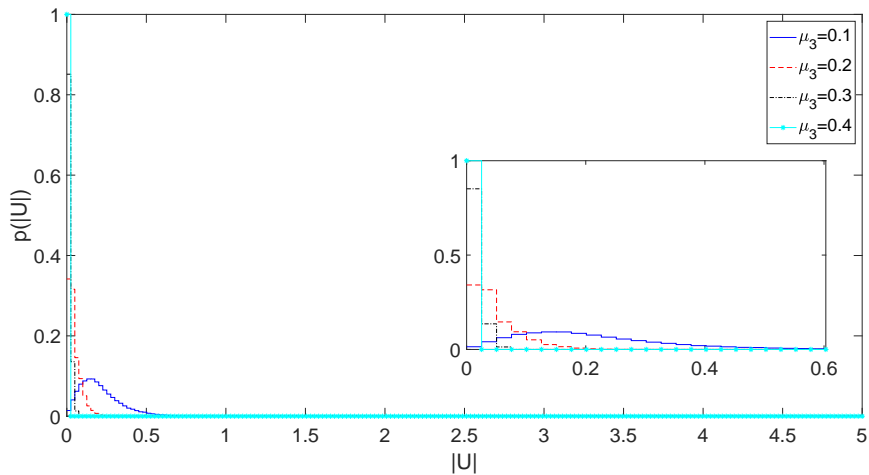}
  \end{center}
\caption{\small Amplitude probability distribution in a chaotic wave field for $m=4, \beta=0.5, \mu_1=1, \mu_2=2, \mu_4 = 2/3$ for various values of $\mu_3$.}
  \label{fig7}
		\end{figure}
\begin{figure}[htb!]
\begin{center}
	\includegraphics[width=0.9\textwidth]{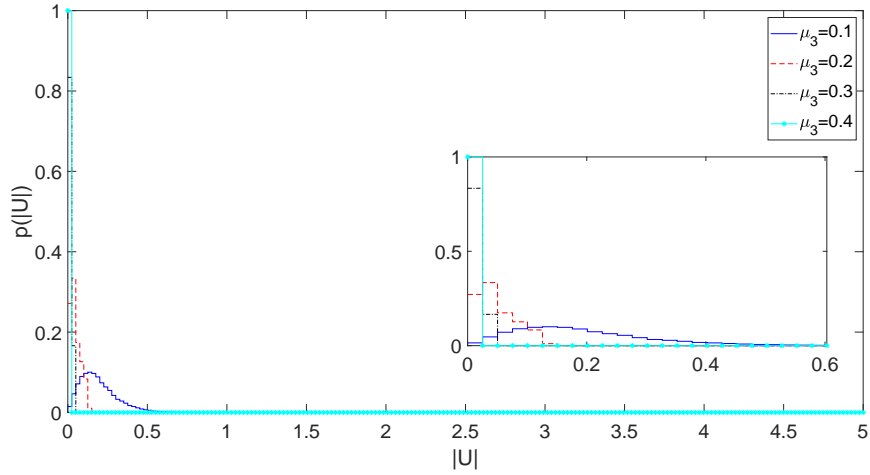}
  \end{center}
\caption{\small Amplitude probability distribution in a chaotic wave field for $m=16, \beta=0.1, \mu_1=1, \mu_2=2, \mu_4 = 2/3$ for various values of $\mu_3$. }
  \label{fig8}
\end{figure}
\begin{figure}[htb!]
\begin{center}
	\includegraphics[width=0.9\textwidth]{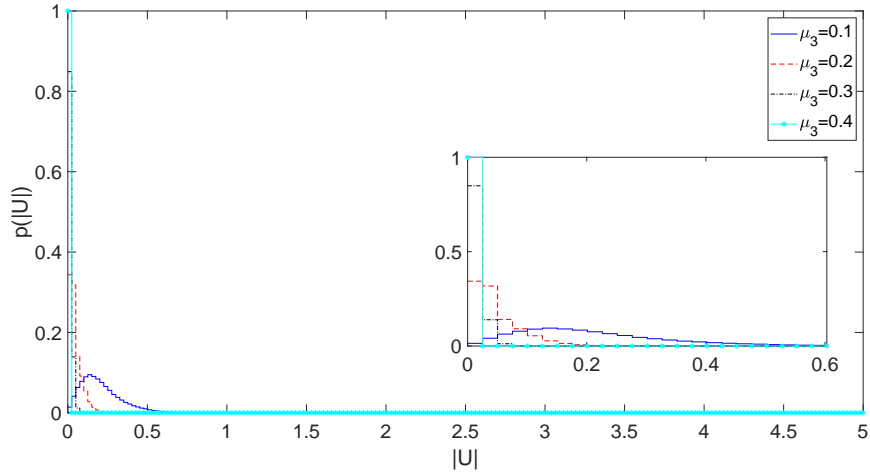}
  \end{center}
\caption{\small Amplitude probability distribution in a chaotic wave field for $m=16, \beta=0.5, \mu_1=1, \mu_2=2, \mu_4 = 2/3$ for various values of $\mu_3$.}
  \label{fig9}
\end{figure}
\begin{figure}[htb!]
\begin{center}
	\includegraphics[width=0.9\textwidth]{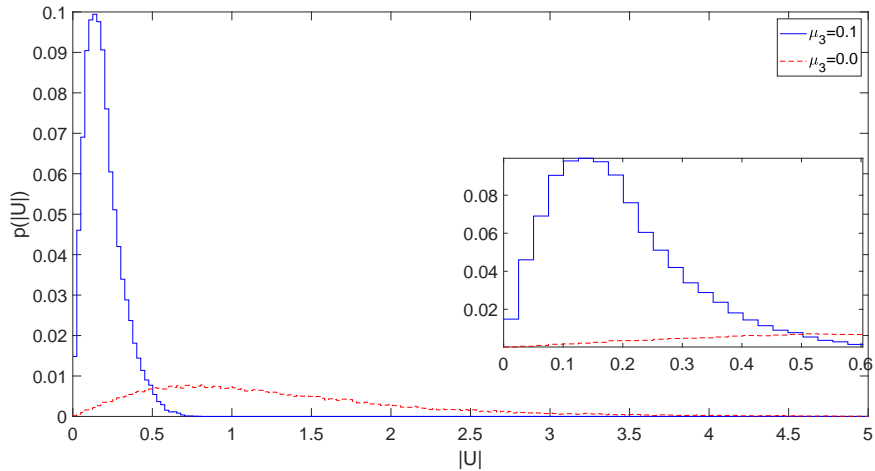}
  \end{center}
\caption{\small Amplitude probability distribution in a chaotic wave field for $m=4, \beta=0.1, \mu_1=1, \mu_2=2, \mu_4 = 2/3$ for $\mu_3 = 0.1$ and $\mu_3=0$.}
  \label{fig10}
\end{figure}

In Fig.~(\ref{fig6}), we plot the amplitude probability distribution in a chaotic wave field for various values of $\mu_3$ using $m=4, \beta=0.1, \mu_1=1, \mu_2=2, \mu_4 = 2/3$. Each of the probability distributions depicted in Figs.~(\ref{fig6})-(\ref{fig10}) include approximately $10^5$ wave components and are recorded after a dimensionless adjustment time of $t=5$ until to the dimensionless time of $t=10$ at various time steps. Checking this figure, one can realize that even the dissipation constant of $\mu_3=0.1$ is strong enough to dissipate rogue waves in the chaotic wave field.
In order to illustrate the effect of the parameter $\beta$ on rogue wave formation probability, we depict Fig.~(\ref{fig7}) using the parameters as $m=4, \beta=0.5, \mu_1=1, \mu_2=2, \mu_4 = 2/3$. It is known that a higher value of $\beta$ leads to an increase in the rogue wave formation probability \cite{BayPRE1}. Comparing Fig.~(\ref{fig6}) and Fig.~(\ref{fig7}), one can realize that the same amount of increase in the dissipation parameter, $\mu_3$, has a more dominant effect than an increase in MI parameter $\beta$.

Additionally, it is also known that an increase in $m$ leads to an increase in the probability of rogue wave formation \cite{BayPRE1}. However, checking Fig.~(\ref{fig8}), it is possible to argue that the effect of dissipation constant is again more significant compared to the MI parameter $m$. The results depicted in Fig.~(\ref{fig8}) are computed using $m=16, \beta=0.1, \mu_1=1, \mu_2=2, \mu_4 = 2/3$.

In order to check the combined effect of increasing both of the MI parameters $\beta$ and $m$, we depict Fig.~(\ref{fig9}) for which the parameters of computations are selected as $m=16, \beta=0.5, \mu_1=1, \mu_2=2, \mu_4 = 2/3$. Although an increase in both of the $\beta$ and $m$ lead to increases in the probability of rogue wave formation, the effect of dissipation coefficient is still more significant than the combined effect of $\beta$ and $m$.

Lastly, we compare the effect of turning the dissipation parameter off. Setting $\mu_3=0$ turns the dKEE into KEE. As shown in Fig.~(\ref{fig10}), MI triggers generation of rogue waves in chaotic wave fields for both of the dKEE and KEE. The value of $\mu_3 = 0.1$ dissipates all the rogue waves, which would exist in the chaotic wave field with no dissipation. With dissipation, the probability distribution of rogue wave amplitudes follows the Rayleigh distribution more closely, however with no dissipation, deviation from the Rayleigh distribution can be observed and the wave amplitude distribution tends to Tayfun distribution. It is possible to state that, the dissipation has a very significant effect on the rogue wave formation compared to the other MI parameters. Similar significant effect would be observed for the gain as well, which could be modeled by using negative dissipation values. 

\section{Conclusion}
In this study, we have studied the effects of dissipation on the probabilities of rogue wave occurrences in the frame of the dissipative Kundu-Eckhaus equation. With this motivation, we have developed a split-step Fourier solver for the numerical solution of the dissipative Kundu-Eckhaus equation and we tested the accuracy and stability of the scheme using an analytical solution. Additionally, we have showed that the MI triggers the generation of chaotic wave fields. We have discussed the effects of various MI parameters and the dissipation coefficient and showed that the probability of rogue wave formation can significantly depend and be controlled by changing the dissipation coefficient. Our results can be used to model the effect of dissipation/gain and damping on rogue wave formation probabilities in various systems. Possible application areas include but are not limited to dissipative optical media, dissipative hydrodynamic media such as the ocean exposed to oil spill and dissipative media in matter physics and Bose-Einstein condensation.

\end{document}